\documentstyle[12pt]{article}
\newcommand{\nc}{\newcommand}
\nc{\al}{\alpha}
\nc{\ald}{{\dot \al}}
\nc{\ba}{\beta_\al}
\nc{\bb}{\beta_\beta}
\nc{\ga}{\g^\al}
\nc{\gb}{\g^\beta}
\nc{\db}{\pa_\beta}
\nc{\dtb}{\delta_\theta^\beta}
\nc{\dab}{{\delta_\al}^\beta}
\nc{\vmab}{V_{-\al}^\beta}
\nc{\vab}{V_\al^\beta}
\nc{\vib}{V_i^\beta}
\nc{\g}{\gamma}
\nc{\G}{\Gamma}
\nc{\D}{\Delta}
\nc{\paj}{P_{-\al}^j}
\nc{\la}{\lambda}
\nc{\La}{\Lambda}
\nc{\var}{\varphi}
\nc{\kvt}{\sqrt{t}}
\nc{\hn}{h^\vee}
\nc{\kn}{k^\vee}
\nc{\pa}{\partial}
\nc{\nn}{\nonumber \\ }
\nc{\hf}{\frac{1}{2}}         
\nc{\dz}{\frac{dz}{2\pi i}}
\nc{\fabc}{{f_{ab}}^c}
\nc{\binomial}[2]{\left (\begin{array}{c} {#1}\\ {#2} \end{array}
\right )}
\nc{\ben}{\begin{equation}}
\nc{\een}{\end{equation}}
\nc{\bea}{\begin{eqnarray}}
\nc{\eea}{\end{eqnarray}}
\nc{\bra}[1]{\langle {#1}|}
\nc{\ket}[1]{|{#1}\rangle}

\nc{\C}{\mbox{\hspace{1.24mm}\rule{0.2mm}{2.5mm}\hspace{-2.7mm} C}}
\nc{\Nat}{\mbox{\hspace{.04mm}\rule{0.2mm}{2.8mm}\hspace{-1.5mm} N}}

\nc{\spa}{\hspace{1 cm},\hspace{1 cm}}
\nc{\vs}{\vspace}
\nc{\NP}[1]{Nucl.\ Phys.\ {\bf #1}}
\nc{\PL}[1]{Phys.\ Lett.\ {\bf #1}}
\nc{\CMP}[1]{Commun.\ Math.\ Phys.\ {\bf #1}}
\nc{\PR}[1]{Phys.\ Rev.\ {\bf #1}}
\nc{\PRL}[1]{Phys.\ Rev.\ Lett.\ {\bf #1}}
\nc{\PTP}[1]{Prog.\ Theor.\ Phys.\ {\bf #1}}
\nc{\PTPS}[1]{Prog.\ Theor.\ Phys.\ Suppl.\ {\bf #1}}
\nc{\MPL}[1]{Mod.\ Phys.\ Lett.\ {\bf #1}}
\nc{\IJMP}[1]{Int.\ Jour.\ Mod.\ Phys.\ {\bf #1}}
\nc{\IM}[1]{Invent.\ Math.\ {\bf #1}}
\nc{\SJNP}[1]{Sov. J. Nucl. Phys.\ {\bf #1}}

\begin{document}

\topmargin -5mm
\oddsidemargin 5mm

\begin{titlepage}
\setcounter{page}{0}
\begin{flushright}
March 1998
\end{flushright}

\vs{8mm}
\begin{center}
{\Large Two-point Functions in Affine Current Algebra}\\[.2cm]
{\Large and Conjugate Weights}

\vs{8mm}
{\large J{\o}rgen Rasmussen}\footnote{e-mail address: 
jorgen@celfi.phys.univ-tours.fr}\\[.2cm]
{\em Laboratoire de Math\'{e}matiques et Physique Th\'{e}orique,}\\
{\em Universit\'{e} de Tours, Parc de Grandmont, F-37200 Tours, France}

\end{center}

\vs{8mm}
\centerline{{\bf{Abstract}}}
\noindent
The two-point functions in 
affine current algebras based on simple Lie algebras
are constructed for all representations, integrable or non-integrable.
The weight of the conjugate field to a primary field of arbitrary weight
is immediately read off.\\[.4cm]

\end{titlepage}
\newpage
\renewcommand{\thefootnote}{\arabic{footnote}}
\setcounter{footnote}{0}

\section{Introduction}

Two-point functions are the simplest non-trivial correlators one may
consider in (extended) conformal field theory. Nevertheless, results in
the case of general representations of affine current algebras
are still lacking, except for $SL(2)$ where invariance under
(loop) projective transformations immediately produces the result.
However, very recently we have provided the general solution in the
case of affine $SL(N)$ current algebra using the differential operator 
realization of simple Lie algebras provided in \cite{PRY4} and explicit
realizations of fundamental representations in terms of fermionic 
creation and annihilation operators \cite{Ras1}.

The objective of the present work is to construct general 
two-point functions in affine current algebras based on any simple
Lie algebra and for all representations, integrable or non-integrable.
Again the construction is
based on the differential operator realization of simple Lie algebras
provided in \cite{PRY4}, in addition to well-known results for general 
fundamental representations and their conjugate (or contragredient)
representations.

Besides providing us with new insight in the general structure of
conformal field theory based on affine current algebra, a motivation for 
studying two-point functions in affine current algebra is found in the
wish to understand how to generalize to higher groups
the proposal by Furlan, Ganchev, Paunov and Petkova
\cite{FGPP} for how Hamiltonian reduction of affine $SL(2)$
current algebra works at the level of correlators. A simple proof
of the proposal in that case
is presented in \cite{PRY2} based on the work \cite{PRY1}
on correlators for degenerate (in particular admissible)
representations in affine $SL(2)$ current algebra. Explicit knowledge
on two-point functions may be seen as a first step in the direction
of understanding that generalization.

Furthermore, an immediate application of knowing the two-point functions
is to determine the weight
of the conjugate (primary) field to a primary field of an arbitrary weight. 
This result is valuable since conjugate 
representations play important roles in various respects, see e.g. 
\cite{Fuc}.
For non-integrable representations such weights are in general not known.

The remaining part of this presentation is organized as follows. In Section 
2 we review our differential operator realization \cite{PRY4} while
fixing the notation. In Section 3 the construction of two-point
functions is provided and the conjugate weights are derived. 
Section 4 contains some concluding remarks, whereas an
illustration is given in Appendix A where we consider elements of
the simple Lie algebra $G_2$.

\section{Notation}

Let {\bf g} be a simple Lie algebra of rank $r$.
{\bf h} is a Cartan subalgebra of {\bf g}. The set of (positive) roots
is denoted ($\Delta_+$) $\Delta$ and the simple roots are 
written $\al_i,\ i=1,...,r$. $\al^\vee = 2\al/\al^2$ is the root 
dual to $\al$. Using the triangular decomposition 
\ben
 \mbox{{\bf g}}=\mbox{{\bf g}}_-\oplus\mbox{{\bf h}}\oplus\mbox{{\bf g}}_+
\een
the raising and lowering operators are denoted $e_\al\in$ {\bf g}$_+$ and
$f_\al\in$ {\bf g}$_-$, respectively, with $\al\in\Delta_+$, and 
$h_i\in$ {\bf h} are the Cartan operators. 
In the Cartan-Weyl basis we have
\ben
 [h_i,e_\al]=(\al_i^\vee,\al)e_\al\spa [h_i,f_\al]=
  -(\al_i^\vee,\al)f_\al
\label{CW}
\een
and
\ben
 \left[e_\al,f_\al\right]=h_\al=G^{ij}(\al_i^\vee,\al^\vee)h_j
\een
where the metric $G_{ij}$ is related to the Cartan matrix $A_{ij}$ as
\ben
 A_{ij}=\al_i^\vee\cdot\al_j=(\al_i^\vee,\al_j)=
  G_{ij}\al_j^2/2
\een
The Dynkin labels $\La_k$ of the weight $\La$ are defined by
\ben
 \La=\La_k\La^{k}\spa \La_k=(\al_k^\vee,\Lambda)
\label{Dynkin}
\een
where $\left\{\La^{k}\right\}_{k=1,...,r}$ is the set of fundamental
weights satisfying
\ben
 (\al_i^\vee,\La^{k})=\delta_i^k
\een 
Elements in $\mbox{\bf g}_+$ may be parameterized using ``triangular 
coordinates" denoted by $x^\al$, one for each positive root, thus we 
write general Lie algebra elements in $\mbox{\bf g}_+$ as
\ben
 g_+(x)=x^\al e_\al \in \mbox{\bf g}_+
\label{gplus}
\een
We will understand ``properly" repeated root indices as in (\ref{gplus})
to be summed over the {\em positive} roots. Repeated Cartan indices as in
(\ref{Dynkin}) are also summed over.
The matrix representation $C(x)$ of $g_+(x)$ in the adjoint 
representation is defined by
\ben
 C_a^b(x)=-x^\beta {f_{\beta a}}^b
\label{cadj}
\een

Now, a differential operator realization $\left\{\tilde{J}_a(x,\pa,\Lambda)
\right\}$ 
of the simple Lie algebra {\bf g} generated by $\left\{j_a\right\}$
is found to be \cite{PRY4}
\bea
\tilde{E}_\al(x,\pa)&=&\vab(x)\db\nn
\tilde{H}_i(x,\pa,\Lambda)&=&\vib(x)\db+\Lambda_i\nn
\tilde{F}_\al(x,\pa,\Lambda)&=&\vmab(x)\db+\paj(x)\Lambda_j
\label{defVP}
\eea
where 
\bea
 \vab(x)&=&\left[B(C(x))\right]_\al^\beta\nn
 \vib(x)&=&-\left[C(x)\right]_i^\beta \nn
 \vmab(x)&=&\left[e^{-C(x)}\right]_{-\al}^\g\left[B(-C(x))\right]_\g^\beta\nn
 \paj(x)&=&\left[e^{-C(x)}\right]_{-\al}^j 
\label{VPQ}
\eea
$B$ is the generating function for the Bernoulli numbers
\ben
  B(u)=\frac{u}{e^u-1}=\sum_{n\geq 0}\frac{B_n}{n!}u^n\nn
\label{Ber}
\een
whereas $\pa_\beta$ denotes partial differentiation wrt $x^\beta$.
Closely related to this differential operator realization is the equivalent
one $\left\{J_a(x,\pa,\Lambda)\right\}$ given by
\bea
 E_\al(x,\pa,\La)&=&-\tilde{F}_\al(x,\pa,\La)\nn
 F_\al(x,\pa,\La)&=&-\tilde{E}_\al(x,\pa,\La)\nn
 H_i(x,\pa,\La)&=&-\tilde{H}_i(x,\pa,\La)
\label{tilde}
\eea
The matrix functions (\ref{VPQ}) are defined in terms of universal
power series expansions, valid for any Lie algebra, but ones that truncate 
giving rise to finite polynomials of which the explicit forms depend on the
Lie algebra in question. Details on the truncations and the resulting
polynomials may be found in \cite{PRY4}.

\subsection{Affine Current Algebra}

Associated to a Lie algebra is an affine Lie algebra characterized by the
central extension
$k$, and associated to an affine Lie algebra is an affine current
algebra whose generators are conformal spin one fields and have amongst
themselves the operator product expansion
\ben
 J_a(z)J_b(w)=\frac{\kappa_{ab}k}{(z-w)^2}+\frac{\fabc J_c(w)}{z-w}
\label{JaJb}
\een
where regular terms have been omitted.
$\kappa_{ab}$ and $\fabc$ are the Cartan-Killing form and the 
structure coefficients, respectively, of the underlying Lie algebra.

It is convenient to collect the traditional multiplet of primary fields
in an affine current algebra (which generically is infinite for
non-integrable representations) in a generating function for that
\cite{FGPP,PRY1,PRY4}, namely the primary field 
$\phi_\La(w,x)$ which must satisfy
\bea
 J_a(z)\phi_\La(w,x)&=&\frac{-J_a(x,\pa,\La)}{z-w}\phi_\La(w,x)\nn
 T(z)\phi_\La(w,x)&=&\frac{\Delta(\phi_\La)}{(z-w)^2}\phi_\La(w,x)
  +\frac{1}{z-w}\pa\phi_\La(w,x)
\label{primdef}
\eea
Here $J_a(z)$ and $T(z)$ are the affine currents and the energy-momentum
tensor, respectively, whereas $J_a(x,\pa,\La)$ are the 
differential operator realizations. $\D(\phi_\La)$ denotes the conformal
dimension of $\phi_\La$.
The explicit construction of primary fields for general simple Lie algebra
and arbitrary representation is provided in \cite{PRY4}. 

An affine transformation of a primary field is given by
\bea
 \delta_\epsilon\phi_\La(w,x)&=&\oint_w\frac{dz}{2\pi i}
  \epsilon^a(z)J_a(z)\phi_\La(w,x)\nn
 &=&\left\{\epsilon^{-\al}(w)V_\al^\beta(x)\pa_\beta
  +\epsilon^i(w)\left(V_i^\beta(x)
  \pa_\beta+\La_i\right)\right.\nn
  &+&\left.\epsilon^\al(w)\left(V_{-\al}^\beta(x)\pa_\beta+
  P_{-\al}^i(x)\La_i\right)\right\}\phi_\La(w,x)
\label{Ward}
\eea
and is parameterized by the $d$ ($d$ is the dimension of the
underlying Lie algebra) independent infinitesimal functions $\epsilon^a(z)$. 

\section{Two-point Functions}

Let $W_2(z,w;x,y;\La,\La')$ denote a general two-point function of two 
primary fields $\phi_\La(z,x)$ and $\phi_{\La'}(w,y)$. From the conformal 
Ward identities or projective invariance the well-known conformal
property of the two-point function is found to be
\ben
 W_2(z,w;x,y;\La,\La')=\frac{\delta_{\D(\phi_\La),\D(\phi_{\La'})}}{(z-w)^{
 \D(\phi_\La)+\D(\phi_{\La'})}}W_2(x,y;\La,\La')
\een
The affine Ward identity 
\ben 
 \delta_\epsilon W_2(z,w;x,y;\La,\La')=
 \langle\delta_\epsilon\phi_\La(z,x)\phi_{\La'}(w,y)\rangle
 +\langle\phi_\La(z,x)\delta_\epsilon\phi_{\La'}(w,y)\rangle=0
\een
may be recast (using (\ref{Ward}))
into the following set of $d$ partial differential equations
\ben
 \left(\tilde{J}_a(x,\pa,\La)+\tilde{J}_a(y,\pa,\La')\right)
  W_2(x,y;\La,\La')=0
\label{ddiff}
\een
It is easily verified that only the $2r$ equations for $a=\pm\al_i$
are independent. By induction, this simply
follows from the fact that $\{\tilde{J}_a\}$
is a differential operator realization of a Lie algebra.
It is the general solution to the equations (\ref{ddiff})
that we shall provide below. 

First we review a few basic properties of fundamental representations
and their conjugate representations. 

In every highest weight representation of highest weight $\La$ the weights 
are given
by $\la=\La-\sum\beta$ where $\sum\beta$ is a sum of positive roots or zero.
The depth of $\la$ is then defined as the height of $\sum\beta$. In a finite
dimensional irreducible highest weight module there exists a unique vector
(up to trivial renormalizations) of lowest weight characterized by
having maximal depth. The conjugate representation of such a representation
is a highest weight representation with highest weight $\La^+$ given by
minus the lowest weight of the original one, while in general all weights in 
the conjugate representation are given by minus the ones in the original
representation. The conjugate representation of a fundamental representation
(which is a finite dimensional irreducible highest weight representation
of highest weight a fundamental weight) is again a fundamental
representation. Due to the uniqueness of the conjugate weight
we shall write $\La^{i^+}=(\La^i)^+$. Many fundamental representations
are self-conjugate, see e.g. \cite{Fuc}.

A key property of the Kronecker product of two finite dimensional
irreducible highest weight representations that we shall use, is the result 
that the singlet occurs in the decomposition of the product if and only if
the two highest weights are conjugate, and in that case its multiplicity
is one, see e.g. \cite{Fuc}. In particular, this statement is valid for the
Kronecker product of two fundamental representations. A simple consequence 
of that is that in the Kronecker product $\La^i\times\La^{i^+}$ there exists
a unique linear combination
\ben
 R^i=\sum C_{\mu\nu}\ket{\la^{(i)}}^\mu\otimes\ket{\la^{'(i^+)}}^\nu
\label{Ri}
\een
which is annihilated by the co-product
\ben
 \D(j_a)=j_a\otimes1+1\otimes j_a
\een
for all generators $j_a\in$ {\bf g}. In (\ref{Ri}) $\{\ket{\la^{(i)}}^\mu\}$
is meant to be a basis for the highest weight module with highest weight
$\La^i$ whereas $\{\ket{\la^{'(i^+)}}^\nu\}$ is a basis for the conjugate
module characterized by $\La^{i^+}$. Let us emphasize that $C_{\mu\nu}$
are uniquely given coefficients (up to an overall and in this respect
irrelevant scaling factor) as soon as the two bases have been chosen. 
Note that $\mu$ and $\nu$ are
multiple indices also counting multiplicities of the weights. This is
illustrated in Appendix A where the self-conjugate fundamental representation
$\La^2$ of $G_2$ is considered.

Hence, in the framework of our differential operator realization 
there exists a unique polynomial $R^i(x,y)$
(again up to scaling) for all $i=1,...,r$ satisfying
\ben
 \left(\tilde{J}_a(x,\pa,\La^i)+\tilde{J}_a(y,\pa,\La^{i^+})\right)
  R^i(x,y)=0
\label{Rconst}
\een
This polynomial may then be decomposed as in (\ref{Ri})
\ben
 R^i(x,y)=\sum C_{\mu\nu}b(x,\La^i,\{j_l\}^\mu)b(y,\La^{i^+},\{j'_{l'}\}^\nu)
\label{Rb}
\een
with the same coefficients $C_{\mu\nu}$. The relation between the 
realizations is
\bea
 \ket{\la^{(i)}}^\mu&=&f_{\al_{j_1}}...f_{\al_{j_{n(\mu)}}}\ket{\La^i}\nn
 b(x,\La^i,\{j_l\}^\mu)&=&\tilde{F}_{\al_{j_1}}(x,\pa,\La^i)...
  \tilde{F}_{\al_{j_{n(\mu)}}}(x,\pa,\La^i)1\nn
  &=&\left(V_{-\al_{j_1}}^\beta(x)\pa_\beta+P_{-\al_{j_1}}^i(x)\right)...\nn
   &\cdot&\left(V_{-\al_{j_{n(\mu)-1}}}^\beta(x)\pa_\beta+
    P_{-\al_{j_{n(\mu)-1}}}^i(x)\right)P_{-\al_{j_{n(\mu)}}}^i(x)
\eea
By construction we have $\al_{j_{n(\mu)}}=\al_i$ for $n(\mu)>0$.

We are now in a position to state our main result:\\[.2cm]
\noindent{\bf Proposition}\\
The two-point function of the primary fields $\phi_\La(z,x)$ and
$\phi_{\La'}(w,y)$ in an affine current algebra is (up to an
irrelevant normalization constant) given by
\ben
 W_2(z,w;x,y;\La,\La')=\frac{\delta_{\D(\phi_\La),\D(\phi_{\La'})}}{(z-w)^{
 \D(\phi_\La)+\D(\phi_{\La'})}}
 \prod_{i=1}^r\left(R^i(x,y)\right)^{p_i(\La,\La')}
\een
where
\bea
 p_i(\La,\La')&=&\La_i\nn
  &=&\La'_{i^+}
\label{mu}
\eea
and $R^i(x,y)$ is given by (\ref{Rb}).\\[.2cm]
{\bf Proof}\\
As remarked above, we only need to consider the actions of 
$\tilde{E}_{\al_j}(x,\pa)+\tilde{E}_{\al_j}(y,\pa)$ and 
$\tilde{F}_{\al_j}(x,\pa,\La)+\tilde{F}_{\al_j}(y,\pa,\La')$ for $j=1,...,r$.
That the $r$ former operators respect (\ref{ddiff}) follows 
directly from (\ref{Rconst}). From (\ref{Rconst}) we also have 
\ben
 \left(V_{-\al}^\beta(x)\pa_{x^\beta}+V_{-\al}^\beta(y)\pa_{y^\beta}\right)
  R^i(x,y)=-\left(P_{-\al}^i(x)+P_{-\al}^{i^+}(y)\right)R^i(x,y)
\een
This implies that
\bea
 &&\left(\tilde{F}_{\al_j}(x,\pa,\La)+\tilde{F}_{\al_j}(y,\pa,\La')\right)
   W_2(x,y;\La,\La')\nn
  &=&\left(\sum_{i=1}^r(\La_i-p_i(\La,\La'))P_{-\al_j}^i(x)
   +\sum_{i=1}^r(\La_{i^+}'-p_i(\La,\La'))
   P_{-\al_j}^{i^+}(y)\right)W_2(x,y;\La,\La')
\eea
from which we obtain (\ref{mu}).\\
$\Box$\\
{}From the condition on the pair $(\La,\La')$ in (\ref{mu}) it follows
immediately that the conjugate weight $\La^+$ to an arbitrary weight
$\La=\sum_{k=1}^r\La_k\La^k$, integrable or non-integrable, is given by
\ben
 \La^+=\sum_{k=1}^{r}\La_k^+\La^k=\sum_{k=1}^{r}\La_{k^+}\La^k
\label{conjla}
\een

\section{Conclusion}

By constructing solutions to the affine Ward identities, we have provided
a general expression for two-point functions
in conformal field theory based on affine current algebra for all simple
Lie groups and all representations, integrable or non-integrable. The
construction relies on the unique existence of a singlet in the Kronecker
product of a fundamental representation and its conjugate
(fundamental) representation. An immediate application is the derivation of 
the conjugate weight to an arbitrary weight.

We hope to come back elsewhere with a discussion on two-point
functions (and conjugate weights) in affine current superalgebra.
In that case one
may employ the recently obtained differential operator realizations 
of the underlying Lie superalgebras \cite{Ras2}.\\[.3cm]
{\bf Acknowledgment}\\[.2cm]
The author thanks Jens Schnittger for fruitful discussions and gratefully 
acknowledges the financial support from the Danish Natural Science
Research Council, contract no. 9700517.

\appendix
\section{The Singlet in $\La^2\times\La^2$ in the Case of $G_2$}

Our labeling of the $r=2$ simple roots in $G_2$ results in the Cartan matrix
\ben
 A=\left(\begin{array}{rr} 2&-3\\  -1&2  \end{array} \right)
\een
This means that in the basis of fundamental weights the simple
roots are given by
\ben
 \al_1=(2,-1)\spa\al_2=(-3,2)
\een
The weights in the 14 dimensional and self-conjugate fundamental
representation $\La^2$ are easily obtained and all have multiplicity 1 except
the weight $(0,0)$ which has multiplicity 2. A particular basis in the
highest weight module $\La^2=(0,1)$ is
\bea
 \ket{0,1}&&\nn
 \ket{3,-1}&=&f_2\ket{0,1}\nn
 \ket{1,0}&=&f_1\ket{3,-1}\nn
 \ket{-1,1}&=&f_1\ket{1,0}\nn
 \ket{-3,2}&=&f_1\ket{-1,1}\nn
 \ket{2,-1}&=&f_2\ket{-1,1}\nn
 \ket{0,0}_1&=&f_1\ket{2,-1}\nn
 \ket{0,0}_2&=&f_2\ket{-3,2}\nn
 \ket{3,-2}&=&f_2\ket{0,0}_2\nn
 \ket{-2,1}&=&f_1\ket{0,0}_1\nn
 \ket{1,-1}&=&f_1\ket{3,-2}\nn
 \ket{-1,0}&=&f_1\ket{1,-1}\nn
 \ket{-3,1}&=&f_1\ket{-1,0}\nn
 \ket{0,-1}&=&f_2\ket{-3,1}
\eea
Here we have introduced the abbreviation $f_i=f_{\al_i}$. Note that
$\ket{0,0}_1$ and $\ket{0,0}_2$ are linearly independent. The singlet linear
combination $R^2$ is worked out to be
\bea
 R^2&=&\ket{0,1}\otimes\ket{0,-1}-\ket{3,-1}\otimes\ket{-3,1}+
  \ket{1,0}\otimes\ket{-1,0}-\ket{-1,1}\otimes\ket{1,-1}\nn
  &+&\left(\ket{-3,2}\otimes\ket{3,-2}+3\ket{2,-1}\otimes\ket{-2,1}\right)\nn
  &-&\left(12\ket{0,0}_1\otimes\ket{0,0}_1
  -6\ket{0,0}_1\otimes\ket{0,0}_2-6\ket{0,0}_2\otimes\ket{0,0}_1
  +4\ket{0,0}_2\otimes\ket{0,0}_2\right)\nn
  &+&\left(\ket{3,-2}\otimes\ket{-3,2}+3\ket{-2,1}\otimes\ket{2,-1}\right)\nn
  &-&\ket{1,-1}\otimes\ket{-1,1}+\ket{-1,0}\otimes\ket{1,0}
   -\ket{-3,1}\otimes\ket{3,-1}+\ket{0,-1}\otimes\ket{0,1}
\eea


\begin{thebibliography}{99}

\bibitem{PRY4} J. Rasmussen, {\em Applications of Free Fields in 2D Current 
 Algebra}, Ph.D. thesis (The Niels Bohr Institute), hep-th/9610167;\\
 J.L. Petersen, J. Rasmussen and M. Yu, \NP{B 502} (1997) 649
\bibitem{Ras1} J. Rasmussen, {\em Two-point Functions in Affine $SL(N)$
 Current Algebra},\\ hep-th/9803114
\bibitem{FGPP} P. Furlan, A.Ch. Ganchev, R. Paunov and V.B. Petkova,
 \PL{B 267} (1991) 63;\\
 P. Furlan, A.Ch. Ganchev, R. Paunov and V.B. Petkova,
 \NP{B 394} (1993) 665;\\
 A.Ch. Ganchev and V.B. Petkova, \PL{B 293} (1992) 56
\bibitem{PRY2}
 J.L. Petersen, J. Rasmussen and M. Yu, \NP{B 457} (1995) 343
\bibitem{PRY1} J.L. Petersen, J. Rasmussen and M. Yu, 
 \NP{B 457} (1995) 309
\bibitem{Fuc} J. Fuchs, {\em Affine Lie Algebras and Quantum Groups},
 (Cambridge Univ. Press, 1992)
\bibitem{Ras2} J. Rasmussen, \NP{B 510} (1998) 688

\end{thebibliography}
\end{document}